\documentclass[10pt,twocolumn,letterpaper]{article}

\usepackage[pagenumbers]{cvpr} 

\definecolor{cvprblue}{rgb}{0.21,0.49,0.74}
\usepackage[pagebackref,breaklinks,colorlinks,allcolors=cvprblue]{hyperref}

\newcommand{\OURS}{TreeMeshGPT}

\newcommand{\boldparagraph}[1]{\vspace{0.1cm}\noindent{\bf #1:} }

\newcommand\www{0.120}
\newcommand{\centered}[1]{\begin{tabular}{l} #1 \end{tabular}}

\usepackage{amsmath,amsfonts,bm}

\def\eqref#1{equation~\ref{#1}}

\def\1{\bm{1}}

\DeclareMathAlphabet{\mathsfit}{\encodingdefault}{\sfdefault}{m}{sl}
\SetMathAlphabet{\mathsfit}{bold}{\encodingdefault}{\sfdefault}{bx}{n}

\usepackage{booktabs}

\usepackage{tikz}
\usepackage{xcolor}
\usepackage{xspace}

\usepackage{epsfig}
\usepackage{graphicx}
\usepackage{amsmath}
\usepackage{amssymb}
\usepackage{caption}
\usepackage{multirow}
\usepackage{url}
\usepackage{xcolor}
\usepackage{wrapfig}
\usepackage{algorithm2e}
\usepackage{comment}

\usepackage{array}       
\usepackage{booktabs}    
\usepackage{siunitx}

\def\paperID{14517} 
\def\confName{CVPR}
\def\confYear{2025}

\title{\OURS: Artistic Mesh Generation with Autoregressive Tree Sequencing}

\author{%
Stefan Lionar$^{1,2,3}$\qquad Jiabin Liang$^{1,2,3}$ \qquad Gim Hee Lee$^{3}$ \vspace{8pt}\\
$^{1}$Sea AI Lab \qquad\quad $^{2}$Garena  \qquad\quad $^{3}$National University of Singapore \vspace{8pt} 
}

\begin{document}

\twocolumn[{%
	\renewcommand\twocolumn[1][]{#1}%
        \maketitle
	\begin{center}
        \vspace{-8mm}
		
        \includegraphics[width=1\linewidth]{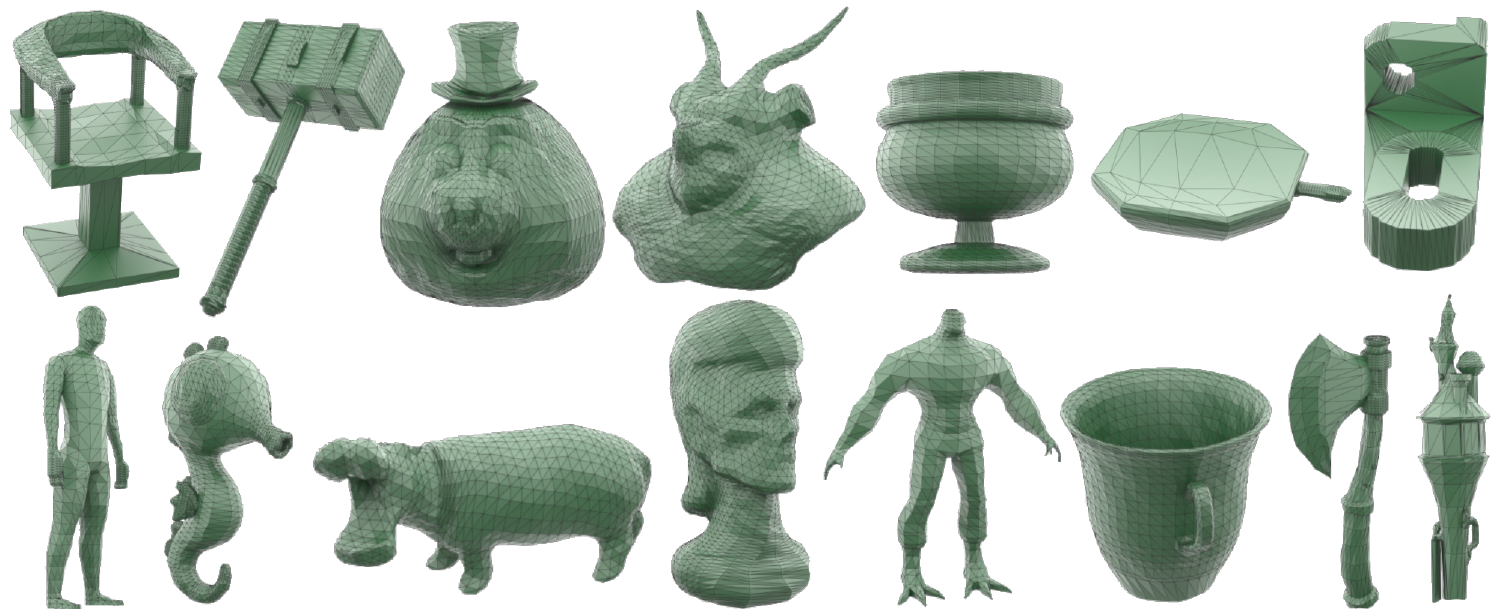}
		\captionsetup{hypcap=false}\captionof{figure}{ \textbf{Artistic meshes generated by \OURS{}.} Our method offers a novel sequencing approach for artistic mesh generation using autoregressive Transformer decoder by retrieving the next token from a dynamically growing tree structure. In our experiment with 7-bit discretization, \OURS{} supports meshes with up to 5,500 triangular faces under strong point cloud conditioning.\protect\footnotemark \protect\footnotemark
        }
		\label{fig:teaser}
	\end{center}    
}]

\footnotetext[1]{\OURS{} supports a higher face count with finer discretization as it preserves the required manifold connectivity condition.}
\footnotetext[2]{Our code is available at: \url{https://github.com/sail-sg/TreeMeshGPT}.}

\maketitle

\begin{abstract}
We introduce TreeMeshGPT, an autoregressive Transformer designed to generate high-quality artistic meshes aligned with input point clouds. Instead of the conventional next-token prediction in autoregressive Transformer, we propose a novel Autoregressive Tree Sequencing where the next input token is retrieved from a dynamically growing tree structure that is built upon the triangle adjacency of faces within the mesh. Our sequencing enables the mesh to extend locally from the last generated triangular face at each step, and therefore reduces training difficulty and improves mesh quality. Our approach represents each triangular face with two tokens, achieving a compression rate of approximately 22\% compared to the naive face tokenization. This efficient tokenization enables our model to generate highly detailed artistic meshes with strong point cloud conditioning, surpassing previous methods in both capacity and fidelity. Furthermore, our method generates mesh with strong normal orientation constraints, minimizing flipped normals commonly encountered in previous methods. Our experiments show that TreeMeshGPT enhances the mesh generation quality with refined details and normal orientation consistency.
\end{abstract}

\section{Introduction}
\label{sec:intro}

In recent advancements in 3D generation, representations such as voxels, point clouds, and implicit functions are often utilized~\cite{li2023generative, xie2022neural}. After the generation process, these representations are converted into meshes using techniques like Marching Cubes~\citep{lorensen1987marching}, which result in dense, over-tessellated triangular meshes. However, these dense meshes are unsuitable for applications that require real-time rendering, such as gaming and virtual reality. Although many mesh down-sampling algorithms can reduce the number of triangles, they also degrade mesh quality and produce messy, unstructured wireframes. In contrast, skilled artists can create highly compact meshes with minimal triangles while preserving the object's sharp details. Additionally, the wireframes created by artists are more regular, aesthetically pleasing, and better aligned with the object's feature or semantic boundaries, which facilitates further human interactions, such as editing and animation. However, the process of manually creating artist-quality meshes is highly time-consuming and labor-intensive.

This laborous process highlights the need for automated methods that can replicate the quality of artist-created meshes without requiring extensive manual effort. MeshAnything~\cite{chen2024meshanything} seeks to bridge the gap between advancements in 3D generation and artist-quality mesh creation by adding point cloud condition to the artistic mesh generation Transformer initially proposed by MeshGPT~\cite{siddiqui2024meshgpt}. Point clouds are chosen as the condition because they are either the direct output or can be obtained conveniently from the generated Marching Cubes meshes of the advanced 3D generation techniques.

MeshAnything~\cite{chen2024meshanything} represents each triangular face with 9 latent tokens, leading to long sequences and limiting artistic mesh generation to 800 faces due to the Transformer’s quadratic complexity. This constraint poses challenges for real-world applications, which often require meshes with significantly higher face counts to accurately represent complex objects and environments. In the subsequent works, MeshAnything V2~\cite{chen2024meshanythingv2} and EdgeRunner~\cite{tang2024edgerunner} leverage triangle adjacency to create shorter sequences to represent the same meshes. Consequently, they are able to generate meshes with up to 1,600 and 4,000 faces, respectively. However, many real-world applications demand meshes with higher face count to accurately represent detailed surface topology. Additionally, challenges remain in generating high-quality meshes free from artifacts such as gaps, missing components, and flipped normals.

To further improve tokenization efficiency and mesh quality, we introduce \OURS{}. Unlike previous methods that rely on conventional next-token prediction in autoregressive Transformers, \OURS{} introduces a novel Autoregressive Tree Sequencing approach. Instead of sequentially predicting next tokens, our method retrieves the next token from a dynamically growing tree structure built upon triangle adjacency within the mesh. This strategy allows the mesh to expand locally from the last generated triangular face at each step, and thus reducing training difficulty and enhancing mesh quality. Our approach represents each triangular face with two tokens, achieving a compression rate of 22\% compared to the naive face tokenization of 9 tokens per face.  This efficient tokenization technique pushes the boundary of artistic mesh generation. With 7-bit discretization, it enables the generation of meshes with up to 5,500 triangles under a strong point cloud condition of 2,048 tokens. Furthermore, our method generates meshes with strong normal orientation constraints, minimizing flipped normals commonly encountered in MeshAnything~\cite{chen2024meshanything} and MeshAnythingV2~\cite{chen2024meshanythingv2}.

In summary, our contributions are as follows:

\begin{itemize}
    \item We propose a novel Autoregressive Tree Sequencing technique that efficiently represents two tokens per triangular face. 
    \item Our proposed tokenization enables the training of a 7-bit discretization artistic mesh generative model with strong point cloud condition, capable of generating high-quality meshes with up to 5,500 faces.
    \item Extensive experiments show that our model can generate higher quality meshes and can generalize to real-world 3D scans.

\end{itemize}

\section{Related Work}
\label{sec:relatedwork}
\subsection{Mesh Extraction}
Constructing a mesh from other 3D representations has been a research focus for decades. Among many successful methods~\cite{bernardini1999ball, kutulakos2000theory, lorensen1987marching}, Marching Cubes~\cite{lorensen1987marching} is the most widely used. It divides a scalar field into cubes and extracts triangles to approximate the isosurface. It is simple yet robust, producing watertight and 2-manifold results. Many improvements such as Dual Contouring~\cite{ju2002dual} and Dual Marching Cubes~\cite{nielson2004dual} have been developed to enhance its capabilities. Another well-known method is Poisson reconstruction~\cite{kazhdan2006poisson}, which uses point clouds and normals as boundary conditions to solve a scalar field defined in 3D space, then applies Marching Cubes to extract the mesh. However, these approaches often focus on representing shapes with dense, over-tessellated meshes, resulting in messy, unstructured wireframes. This makes them unsuitable for downstream workflows that require efficient and structured meshes, such as real-time rendering, animation rigging, and editing. The dense, over-tessellated meshes not only increase computational load but also lack the regularity and semantic alignment necessary for the downstream processes.

\subsection{3D Generation}
After the great success of 2D image generation~\cite{rombach2022high}, 3D generation has become a promising research direction. This field focuses on generating 3D assets for industries such as gaming, film, and AR/VR. Due to the limited availability of 3D data, early methods~\cite{poole2023dreamfusion,jain2022zero, shi2024mvdream, michel2022text2mesh} relied on optimizing underlying representations to mimic conditioning from 2D images or multi-view 2D images.

With the introduction of large-scale datasets~\citep{deitke2023objaverse,deitke2023objaversexl}, feed-forward 3D generation techniques, such as those in~\cite{hong2023lrm, xu2024instantmesh,li2024instant3d}, have become feasible. These techniques significantly improve generation speed compared to optimization-based methods. However, the resulting meshes often suffer from lower quality and lack diversity.

Inspired by the success of 2D diffusion models in image generation, many researchers have attempted to apply diffusion techniques directly to 3D data~\cite{nichol2022point, cheng2023sdfusion, wang2023rodin, zhang2024clay}. For instance, CLAY~\cite{zhang2024clay}, a transformer-based 3D latent diffusion model, achieves state-of-the-art results in high-quality shape generation. Despite these advances, these generation methods often require post-conversion for downstream applications, which remains a non-trivial challenge.

\subsection{Autoregressive Mesh Generation}

To address these limitations,  recent approaches have leveraged autoregressive models for direct mesh generation~\cite{nash2020polygen, siddiqui2024meshgpt, chen2024meshxl,weng2024pivotmesh}. MeshGPT~\citep{siddiqui2024meshgpt} was the first to tokenize a mesh through face sorting and compress it using a VQ-VAE~\citep{van2017neural, lee2022autoregressive}, followed by an autoregressive transformer to predict the compressed token sequence. This method enables the generation of meshes with direct supervision from artist-created topology information, which is often absent in previous approaches.

Subsequent works~\citep{chen2024meshxl, weng2024pivotmesh, chen2024meshanything} explored more efficient representations and incorporated input conditioning, such as point clouds and images. However, these methods are limited to generating meshes with fewer than 800 faces due to the long sequence lengths and the quadratic computational cost of transformers. MeshAnythingV2~\citep{chen2024meshanythingv2} and EdgeRunner~\cite{tang2024edgerunner} introduced more compact mesh tokenization techniques that leverage triangle adjacency, increasing the maximum face count to 1,600 and 4,000, respectively. Meshtron~\citep{hao2024meshtron} proposes hourglass architecture and sliding window inference to scale up face count capacity while using the naive tokenization~\citep{chen2024meshxl}. In a concurrent work, BPT~\citep{weng2024scaling} proposes a compact tokenization using block-wise indexing and localized patch aggregation. These methods rely on the next-token prediction commonly used in large language models (LLMs) and autoregressive image generation. In contrast, we offer a novel sequencing strategy based on a dynamically growing tree structure, aiming to increase the maximum face count and improve the overall quality of the generated meshes.

\section{Method}
\label{sec:method}

\subsection{Autoregressive Tree Sequencing}

\begin{figure*}[t!]
    \centering
    \includegraphics[width=\textwidth]{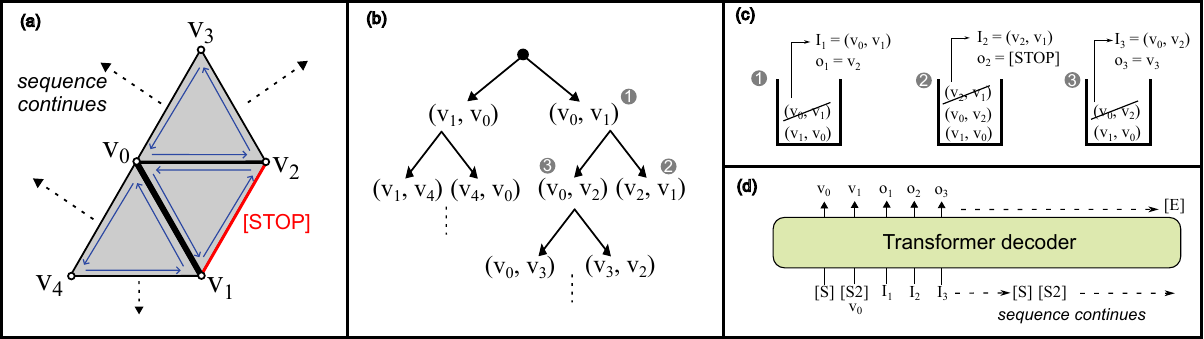}

\caption{\textbf{Illustration of the sequence order in our Autoregressive Tree sequencing}. \textbf{\textit{a). A small subset of a triangular mesh.}} \texttt{[STOP]} indicates boundary.  \textbf{\textit{b). An equivalent tree representation of the mesh.}} In this tree, each node is represented as a directed edge from a pair of vertices. The root is initialized with two child nodes: $(v_0, v_1)$ and its twin $(v_1, v_0)$. A DFS traversal then proceeds to create the input-output sequence.  \textbf{\textit{c). Dynamic stack from the DFS traversal.}} The stack is initialized with $(v_0, v_1)$ and its twin $(v_1, v_0)$. The input $I_n$ is always obtained from the top of the stack. Thus, $I_1 = (v_0, v_1)$ at \textit{step 1}. The opposite vertex of $I_1$ is $v_2$ and consequently, $o_1$ is set to $v_2$. Two new edges are then formed by connecting the opposite vertex to the initial pair of vertices: $(v_2, v_1)$ and $(v_0, v_2)$. The direction is enforced to be counter-clockwise on the potential next adjacent faces. \textit{At step 2,} $I_2 = (v_2, v_1)$. Since $I_2$ is a boundary, $o_2$ is set to \texttt{[STOP]} label and no new edge is added to the stack. \textit{Step 3} and onwards follow the same traversal process. \textbf{\textit{d). Transformer decoder sequence.}} The sequence in the Transformer decoder follows the input-output pairs from the tree traversal. The auxiliary tokens to initialize the generation of a connected component and the \texttt{[EOS]} are also added to the input-output sequence.}  
    \label{fig:illustration} \vspace{-0.2cm}
\end{figure*}

Tokenization plays a crucial role in autoregressive models, particularly in complex tasks like 3D mesh generation, where both the quality and efficiency of tokenization significantly affect model performance and scalability. In the context of mesh generation, tokenization involves encoding vertices, edges, or faces into sequential tokens that the model can process step-by-step. Drawing insights from LLMs, previous autoregressive mesh generation methods follow next-token prediction strategy that explicitly uses each output as the input for the subsequent step. 

Our approach differs from those methods by using a tree-based traversal scheme to grow the mesh during the generation process. Specifically, in \OURS{}, the input to the Transformer decoder are directed mesh edges, represented as \( \mathbf{I} = \{(v^n_1, v^n_2)\}_{n=1}^{N} \in \mathbb{R}^{N \times 6}\), where each \((v^n_1, v^n_2)\) denotes a pair of dequantized vertices for the generation at step $n$. At each step, the Transformer decoder makes a localized prediction to either add a new vertex $v^n_3$ to expand the mesh by connecting to the initial pair of vertices $(v^n_1, v^n_2)$ or predict \texttt{[STOP]} label indicating that no further expansion should occur from the input edge. \OURS{} leverages a tree traversal process to construct the sequential input-output pairs for the autoregressive generation, described as follows (Note: the sequence order description below omits auxiliary tokens for simplicity).

\boldparagraph{Sequence order} We utilize a half-edge data structure with depth-first-search (DFS) traversal to construct the sequential input-output pairs, $(\mathbf{I}, \mathbf{O}) = \{ (I_n, o_n) \}_{n=1}^{N}$, accompanied by a dynamic stack \( \mathbf{S} \) to manage the traversal process.

The traversal process starts from a directed edge in a mesh, in which we determine if this edge has an opposite vertex that forms a triangle, is a boundary, or if the triangle has already been visited. When a new triangle is formed, the output \( o_n \) is defined as the opposite vertex \( v^n_3 \). Two new edges are created by connecting \( v^n_3 \) to the initial edge’s vertices \( I_n = (v^n_1, v^n_2) \), resulting in edges \((v^n_3, v^n_2)\) and \((v^n_1, v^n_3)\). These edges are directed in a counter-clockwise orientation on the potential next adjacent faces, as enforced by the half-edge data structure. The newly created edges are then pushed onto the stack \( \mathbf{S} \) for continued traversal:

\vspace{-0.55cm}

\[
\mathbf{S} := \mathbf{S} \odot (v^n_1, v^n_3) \quad \text{and}  \quad  \mathbf{S} := \mathbf{S} \odot (v^n_3, v^n_2) \vspace{-0.1cm}
\] where \(\odot \) represents the operation of pushing the edge \((v_i, v_j)\) onto the top of stack \( \mathbf{S} \). Conversely, if the input edge is a boundary or if adding a new vertex forms a previously visited triangle, the output \( o_n \) is set to the \texttt{[STOP]} label. In this case, no new vertex or edge is added. 

The input for the next step, \( n+1 \), is obtained by popping the top edge from the stack:

\vspace{-0.55cm}

\[
I_{n+1} = (v^{n+1}_1, v^{n+1}_2) := \text{top}(\mathbf{S}), \quad \mathbf{S} := \mathbf{S} \setminus \text{top}(\mathbf{S}) \vspace{-0.1cm}
\] where \( \text{top}(\mathbf{S}) \) retrieves the current top edge of the stack \( \mathbf{S} \), and \( \mathbf{S} := \mathbf{S} \setminus \text{top}(\mathbf{S}) \) updates \( \mathbf{S} \) by removing this top edge. We initialize the stack with a directed edge at the lowest position of the mesh and its twin. The traversal then proceeds until all triangles are visited. A simple illustration of this sequencing process is provided in Figure~\ref{fig:illustration}.

Note that a mesh may consist of multiple connected components. A component begins from an initial edge, expands as edges are added to a stack, and is considered fully traversed once the stack is empty. When there are multiple components in a mesh, the traversal of the first component starts from the edge with the lowest position in the mesh and continues until no edge remains in the stack. Each subsequent component begins from the next available edge positioned at the lowest among the remaining unvisited faces.

\boldparagraph{Generation process} To initialize the generation of a mesh component, an auxiliary token \texttt{[SOS]} \( \in \mathbb{R}^D \) is used as the input to predict the first vertex \( v_1 \). For the next step, a second auxiliary token \texttt{[SOS2]} \( \in \mathbb{R}^C \), concatenated with the embedded representation of $v_1$, serves as input to predict the second vertex \( v_2 \). Once these initial vertices are predicted, the mesh generation proceeds through our Autoregressive Tree Sequencing with a stack initialized by $(v_1, v_2)$ and its twin $(v_2, v_1)$. When the stack is empty—indicating the current component is complete—a new component is initialized by reintroducing the \texttt{[SOS]} and \texttt{[SOS2]} tokens. After all components have been generated, the sequence is terminated with an \texttt{[EOS]} label.

The final mesh is constructed by gathering the faces formed from the initial input vertex pairs and their predicted opposite vertices: $\mathcal{M} = \bigcup_{n=1}^{N} (v^n_1, v^n_2, v^n_3)$ for the generation steps where $I_n \notin \{\texttt{[SOS]}, \texttt{[SOS2]}\} \text{ and } o_n \notin \{\texttt{[STOP]}, \texttt{[EOS]}\}\}$.

\boldparagraph{Input embedding} We employ a positional embedding function from~\cite{zhang20233dshape2vecset} to encode each vertex into a high-dimensional space, capturing its positional information across multiple frequency bases. This embedding function \( \text{PosEmbed(.)}: \mathbb{R}^3 \rightarrow \mathbb{R}^C \) maps 3D coordinates to a \( C \)-dimensional embedding. For each edge, the embeddings of its vertex pair are concatenated, creating a representation in \( \mathbb{R}^{2C} \), which is subsequently passed through an MLP to map it to the Transformer’s hidden dimension, \( \mathbb{R}^{2C} \rightarrow \mathbb{R}^D \).

\boldparagraph{Vertex prediction} In prior works on mesh generation, each vertex is represented as a sequence of three tokens corresponding to its quantized \( x \)-, \( y \)-, and \( z \)-coordinates. Specifically, to predict a single vertex’s position, those models generates each coordinate independently as separate tokens in sequence. This approach leads to longer sequences, as each vertex requires three distinct tokens. In contrast, our method predicts the vertex' quantized \( x \)-, \( y \)-, and \( z \)-coordinates in one sequence length by using hierarchical MLP heads. This hierarchical approach maintains the sequential nature in predicting the \( x \)-, \( y \)-, and \( z \)-coordinates. Further details can be found in the supplementary material. As shown in the ablation study (Section~\ref{sec:ablation}), our hierarchical MLP heads result in easier coordinate sampling compared to prediction heads that predict the \( x \)-, \( y \)-, and \( z \)-coordinates simultaneously.

\boldparagraph{Advantages} Our Autoregressive Tree Sequencing approach adds only two sequence steps per triangular face as each face introduces two new nodes during the tree traversal process. Additionally, since most meshes consist of only a few connected components, our method requires minimal auxiliary tokens. This efficient sequencing achieves a compression rate of approximately 22\% for most meshes compared to naive tokenization, which uses 9 tokens per face. Our compression rate is thus approximately double that of methods like MeshAnythingV2~\cite{chen2024meshanythingv2} and EdgeRunner~\cite{tang2024edgerunner}. Additionally, by using a dynamic stack to manage the input sequence, our method allows the Transformer to focus solely on making localized predictions at each step, hence improving training efficiency. Furthermore, our method generates mesh with strong normal orientation constraints, minimizing flipped normals commonly encountered in MeshAnything~\cite{chen2024meshanything} and MeshAnythingV2~\cite{chen2024meshanythingv2}.

\boldparagraph{Loss function} We aim to train the Transformer decoder $\theta$ to maximize the likelihood of generating the sequence of outputs \(\{o_n\}_{n=1}^N\) given the input sequence \(\{I_n\}_{n=1}^N\):

\vspace{-0.4cm}
\begin{equation}
    \prod_{n=1}^{N} P(o_n \mid I_{\leq n}; \theta).
\end{equation}
\vspace{-0.3cm}

To this end, given the ground truth and predicted values with teacher-forcing across all steps, denoted by \( \mathbf{O} = \{\mathbf{O}_x, \mathbf{O}_y, \mathbf{O}_z\} \) and \( \hat{\mathbf{O}} = \{\hat{\mathbf{O}}_x, \hat{\mathbf{O}}_y, \hat{\mathbf{O}}_z\} \), respectively, where each \( \mathbf{O}_{.} \) represents the discretized vertex coordinate along a specific axis, we use a loss function defined as the sum of cross-entropy losses for each coordinate:

\vspace{-0.5cm}
\begin{equation}
    \mathcal{L} = \mathcal{L}_{\text{CE}}(\mathbf{O}_x, \hat{\mathbf{O}}_x) + \mathcal{L}_{\text{CE}}(\mathbf{O}_y, \hat{\mathbf{O}}_y) + \mathcal{L}_{\text{CE}}(\mathbf{O}_z, \hat{\mathbf{O}}_z).
\end{equation}
\vspace{-0.5cm}

To incorporate stopping conditions, we add the \texttt{[STOP]} and \texttt{[EOS]} labels to the class selection on the height axis, extending it with two additional classes beyond the discretized coordinate classes.

\section{Experiments}

\subsection{Dataset}

We use Objaverse~\cite{deitke2023objaverse} meshes as our training dataset. To ensure high-quality meshes, we select meshes that meet the half-edge traversal requirement, i.e., they are manifold and have no flipped normal. All meshes are preprocessed by centering and normalizing them within a cube spanning $[-0.5, 0.5]$. We apply 7-bit discretization, remove any duplicate triangles, and choose meshes with less than 5,500 faces. Additionally, we perform orthographic projections and exclude meshes where one of the projections has extremely small area or contains more than one cluster. After filtering, we retain a total of 75,000 meshes, of which 1,000 are reserved for validation, with the remainder used for training. To increase data diversity, we apply the following augmentations:
\begin{itemize}
    \item \textbf{Scaling:} Each axis is scaled independently by a factor randomly chosen from the range \([0.75, 0.95]\).
    \item \textbf{Rotation:} We first apply a \(90^\circ\) or \(-90^\circ\) rotation along the \( x \)- or \( y \)-axis with a probability of $0.3$. Afterward, we perform a rotation around the \( z \)-axis with an angle uniformly sampled from \([-180^\circ, 180^\circ]\).

\end{itemize}

\subsection{Implementation Details}

\boldparagraph{Point cloud conditioning} Following previous approaches~\cite{tang2024edgerunner, chen2024meshanything, chen2024meshanythingv2}, we adopt point cloud conditioning to provide practical guidance for the generation process. To achieve this, we sample a point cloud from the input mesh surface and apply a lightweight encoder~\cite{chen2024meshanything, chen2024meshanythingv2, zhang20233dshape2vecset, zhang2024clay}. Specifically, we sample $8192$ points, denoted as \( \mathbf{X} \in \mathbb{R}^{8192 \times 3} \), from the mesh surface. A cross-attention layer is then used to encode these points into a latent code:

\vspace{-0.2cm}
\begin{equation} 
  \mathbf{Z} = \text{CrossAtt}\big(\mathbf{Q}, \text{PosEmbed}(\mathbf{X})\big) 
\end{equation} where \( \mathbf{Q} \in \mathbb{R}^{2048 \times C} \) represents query embeddings with a shorter sequence length of $2048$ and hidden dimension of \( C \), \( \text{PosEmbed}(.) \) is the same input embedding function in our Autoregressive Tree Sequencing, and \( \mathbf{Z} \in \mathbb{R}^{2048 \times L} \) is the resulting latent code. The latent code \( \mathbf{Z} \) is then prepended to the initial \texttt{[SOS]} token in the Transformer decoder to provide point cloud-based conditioning for mesh generation. 

\boldparagraph{Architecture details} Our model employs a Transformer decoder with 24 layers, 16 attention heads, and a hidden dimension of 1024 and adds the sinusoidal positional encoding~\cite{vaswani2017attention} to encode the token position. We apply a full self-attention for the latent code condition and a causal self-attention mask for the autoregressive decoder. PyTorch's FlexAttention is used to implement this attention mask efficiently. Additionally, we adopt fp16 mixed-precision to optimize computational speed and memory efficiency. We set the hidden dimension $C$ of the positional embedding (PosEmbed) to 512 and use 7-bit quantization to discretize the coordinate output into 128 classes.

\boldparagraph{Training details} We use AdamW~\cite{loshchilov2019decoupled, kingma2015adam} with a learning rate of \( 10^{-4} \), \( \beta_1 = 0.9 \) and \( \beta_2 = 0.99 \) as the optimizer. Our model is trained with $8\times$ A100-80GB GPUs for 5 days with an effective batch size of 128.

\boldparagraph{Sampling strategy} We use a multinomial sampling strategy with a top-$k$ of 5 during the generation process. Empirically, we find that adjusting the temperature at different stages achieves an optimal balance between diversity and generation quality. Specifically, we set the temperature to 1 when the stack length is below 10, reduce it to 0.4 when the stack length is below 30, and further decrease it to 0.2 beyond that.

\boldparagraph{Inference adjustment} We find that a few adjustments during inference can improve the generation performance. First, we keep track of the generated faces for each step. If our model predicts a vertex that would form a triangle that is a duplicate to the previously generated faces, we adjust the prediction to \texttt{[STOP]} and retrieve the next input from the top of the stack. This checking operation is fast since the faces are of discrete tensor.

Additionally, we observe that our model often struggles to predict the \texttt{[EOS]} label in longer sequences. In these cases, while the \texttt{[EOS]} label consistently appears among the top 5 logits, it is rarely sampled. To address this, we apply an addition factor to the logit of \texttt{[EOS]}, incrementally increasing this factor each time \texttt{[EOS]} appears in the top 5 logits. To avoid early \texttt{[EOS]} prediction after this adjustment, we bypass multinomial top-$k$ sampling for \texttt{[EOS]} and select it only when it becomes the highest logit.

\subsection{Results on Objaverse Dataset}

\subsubsection{Qualitative Results}

\begin{figure*}[!h]
\vspace{-2mm}
    \centering
    \setlength{\tabcolsep}{1pt}
    \begin{tabular}{c|cc|cc|cc} 

        \textbf{GT Mesh} & \multicolumn{2}{c|}{\textbf{MeshAnything}~\citep{chen2024meshanything}} &\multicolumn{2}{c|}{\textbf{MeshAnythingV2}~\citep{chen2024meshanythingv2}} & \multicolumn{2}{c}{\textbf{Ours}} \\ 


        \centered{\includegraphics[width=\www\linewidth]{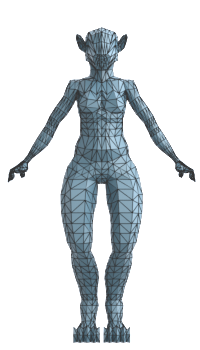}}
        & \centered{\includegraphics[width=\www\linewidth]{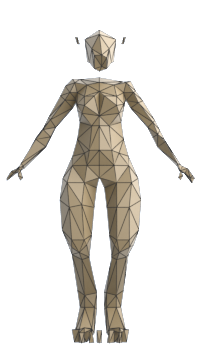}}
        & \centered{\includegraphics[width=\www\linewidth]{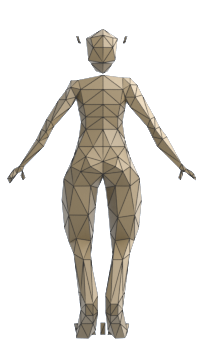}}
         & \centered{\includegraphics[width=\www\linewidth]{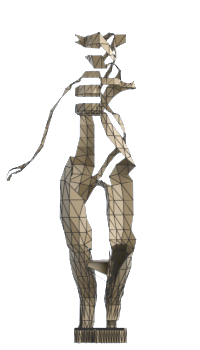}}
          & \centered{\includegraphics[width=\www\linewidth]{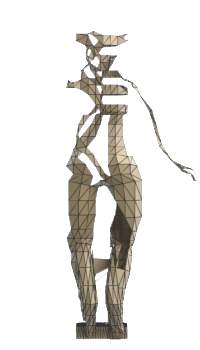}}
           & \centered{\includegraphics[width=\www\linewidth]{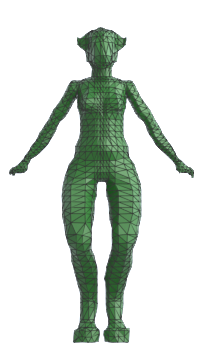}}
            & \centered{\includegraphics[width=\www\linewidth]{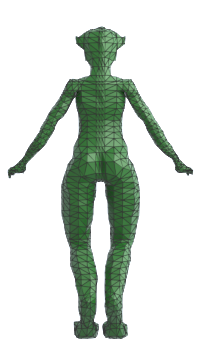}}
         \\

     \#Face = 4130 & \multicolumn{2}{c|}{\#Face = 439} & \multicolumn{2}{c|}{\#Face = 2247} & \multicolumn{2}{c}{\#Face = 3311} \\

    \centered{\includegraphics[width=\www\linewidth]{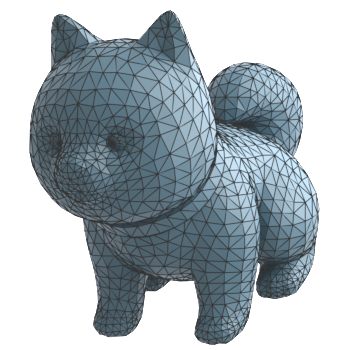}}
        & \centered{\includegraphics[width=\www\linewidth]{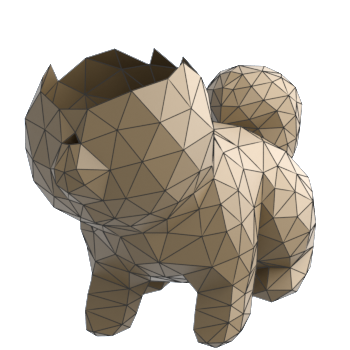}}
        & \centered{\includegraphics[width=\www\linewidth]{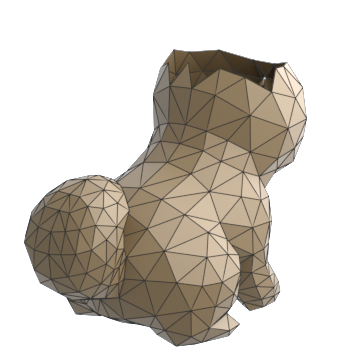}}
         & \centered{\includegraphics[width=\www\linewidth]{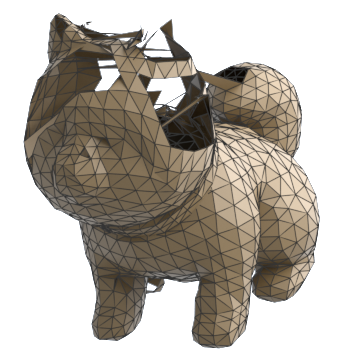}}
          & \centered{\includegraphics[width=\www\linewidth]{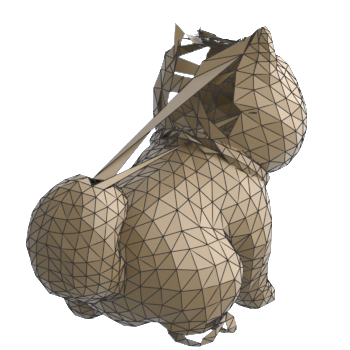}}
           & \centered{\includegraphics[width=\www\linewidth]{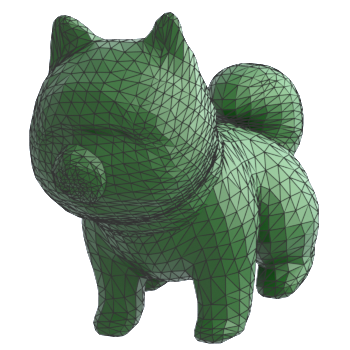}}
            & \centered{\includegraphics[width=\www\linewidth]{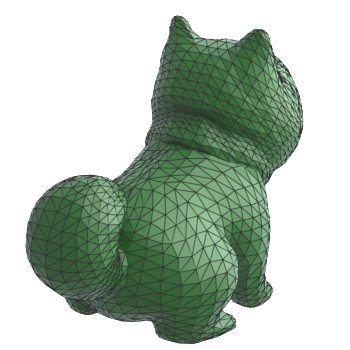}}
            \\
     \#Face = 4316 & \multicolumn{2}{c|}{\#Face = 800} & \multicolumn{2}{c|}{\#Face = 2246} & \multicolumn{2}{c}{\#Face = 4545} \\ 

     \centered{\includegraphics[width=\www\linewidth]{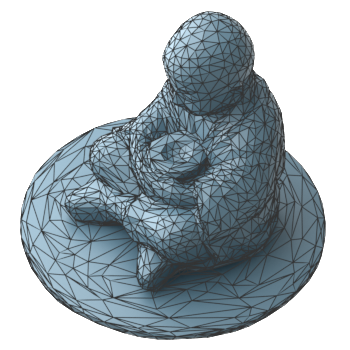}}
        & \centered{\includegraphics[width=\www\linewidth]{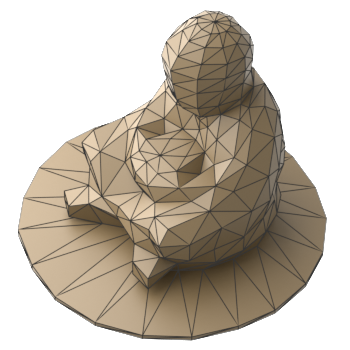}}
        & \centered{\includegraphics[width=\www\linewidth]{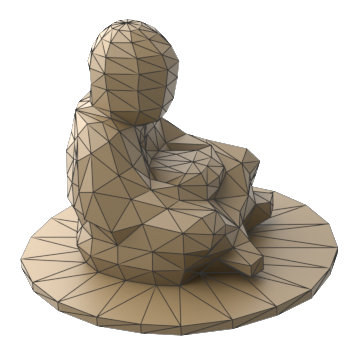}}
         & \centered{\includegraphics[width=\www\linewidth]{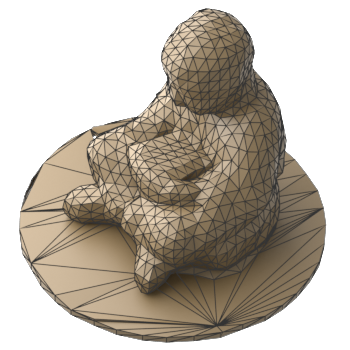}}
          & \centered{\includegraphics[width=\www\linewidth]{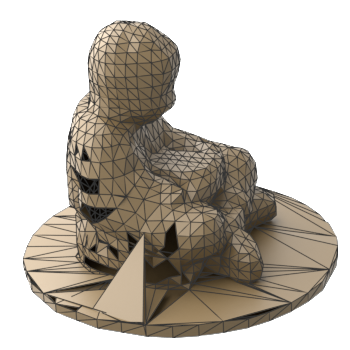}}
           & \centered{\includegraphics[width=\www\linewidth]{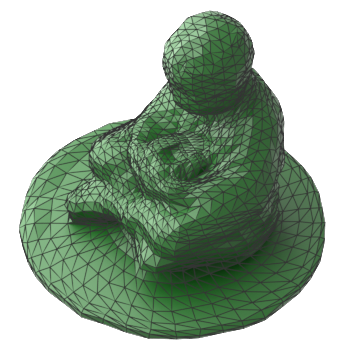}}
            & \centered{\includegraphics[width=\www\linewidth]{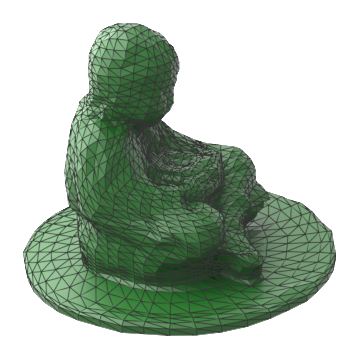}}
        \\

     \#Face = 3718 & \multicolumn{2}{c|}{\#Face = 630} & \multicolumn{2}{c|}{\#Face = 2090} & \multicolumn{2}{c}{\#Face = 5122} \\
    
        \centered{\includegraphics[width=\www\linewidth]{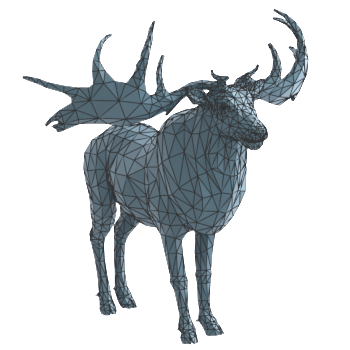}}
        & \centered{\includegraphics[width=\www\linewidth]{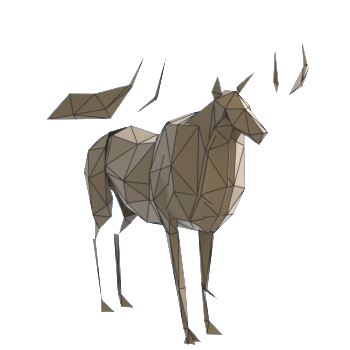}}
        & \centered{\includegraphics[width=\www\linewidth]{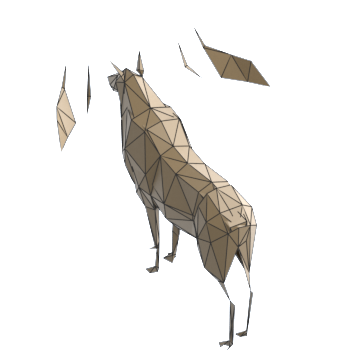}}
         & \centered{\includegraphics[width=\www\linewidth]{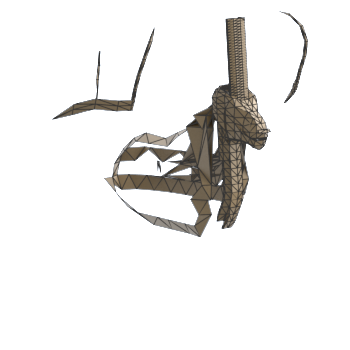}}
          & \centered{\includegraphics[width=\www\linewidth]{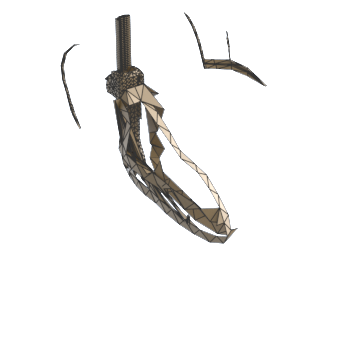}}
           & \centered{\includegraphics[width=\www\linewidth]{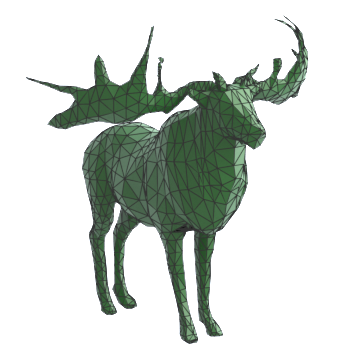}}
            & \centered{\includegraphics[width=\www\linewidth]{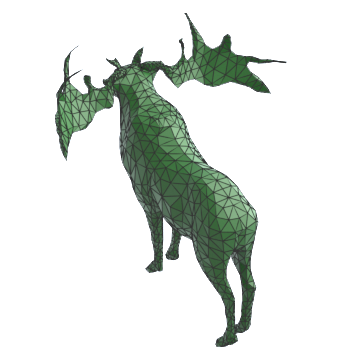}}
        \\

     \#Face = 3694 & \multicolumn{2}{c|}{\#Face = 576} & \multicolumn{2}{c|}{\#Face = 1998} & \multicolumn{2}{c}{\#Face = 2708} \\ 

        
        \centered{\includegraphics[width=\www\linewidth]{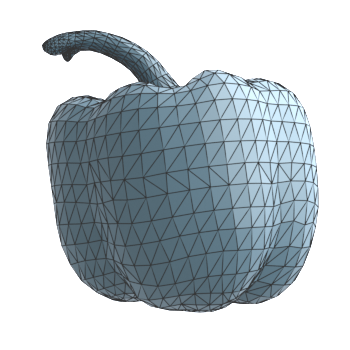}}
        & \centered{\includegraphics[width=\www\linewidth]{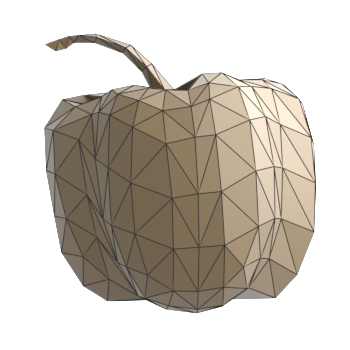}}
        & \centered{\includegraphics[width=\www\linewidth]{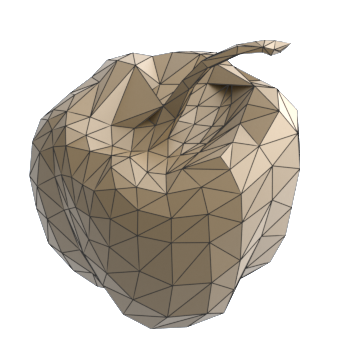}}
        & \centered{\includegraphics[width=\www\linewidth]{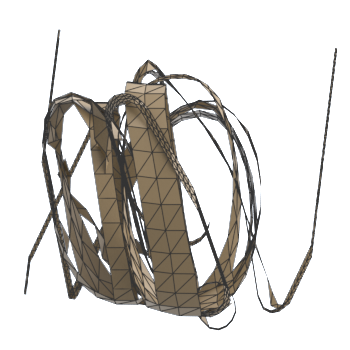}}
        & \centered{\includegraphics[width=\www\linewidth]{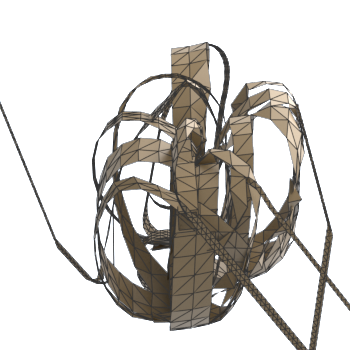}}
     & \centered{\includegraphics[width=\www\linewidth]{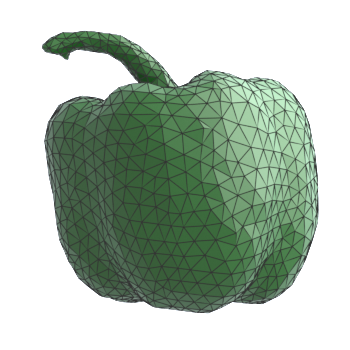}}
           & \centered{\includegraphics[width=\www\linewidth]{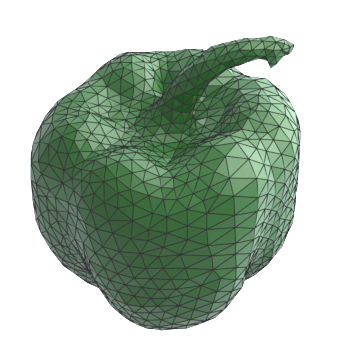}}

        \\

     \#Face = 2442 & \multicolumn{2}{c|}{\#Face = 495} & \multicolumn{2}{c|}{\#Face = 2159} & \multicolumn{2}{c}{\#Face = 2646} \\

        
     \centered{\includegraphics[width=0.21\linewidth]{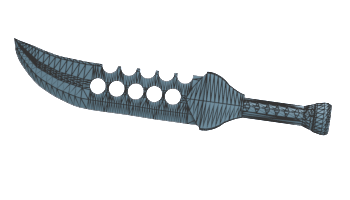}}
        & \multicolumn{2}{c|}{\centered{\includegraphics[width=0.21\linewidth]{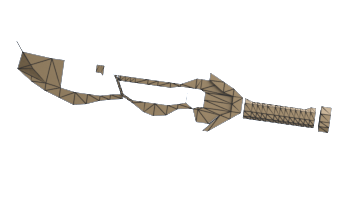}}}
         & \multicolumn{2}{c|}{\centered{\includegraphics[width=0.21\linewidth]{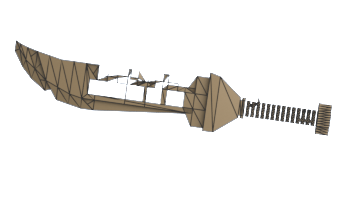}}}
          & \multicolumn{2}{c}{\centered{\includegraphics[width=0.21\linewidth]{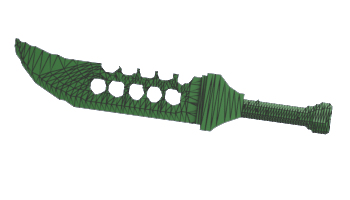}}}
        \\

     \#Face = 3638 & \multicolumn{2}{c|}{\#Face = 584} & \multicolumn{2}{c|}{\#Face = 1769} & \multicolumn{2}{c}{\#Face = 2114} \\

     \centered{\includegraphics[width=0.17\linewidth]{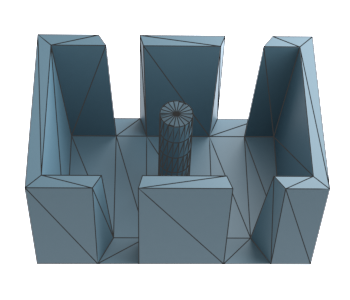}} 
        & \multicolumn{2}{c|}{\centered{\includegraphics[width=0.17\linewidth]{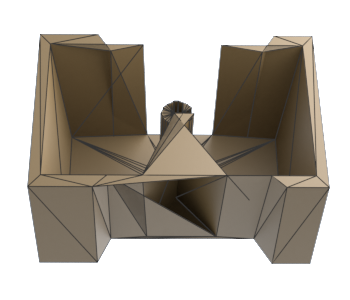}}}
         & \multicolumn{2}{c|}{\centered{\includegraphics[width=0.17\linewidth]{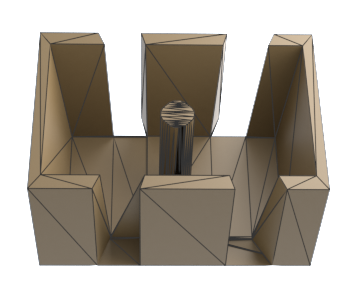}}}
          & \multicolumn{2}{c}{\centered{\includegraphics[width=0.17\linewidth]{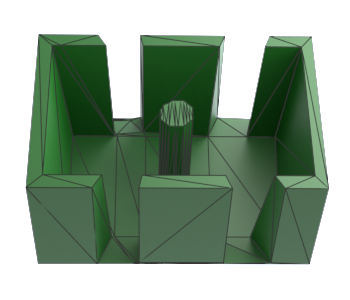}}}
        \\ 
        
     \#Face = 740 & \multicolumn{2}{c|}{\#Face = 235} & \multicolumn{2}{c|}{\#Face = 265} & \multicolumn{2}{c}{\#Face = 301} \\

    \end{tabular}
    \caption{
    \textbf{Qualitative comparison on Objaverse dataset~\citep{deitke2023objaverse}.} Our model is able to generate meshes with higher face counts and refined details compared to the baselines. Results from the baselines use point clouds sampled from marching cube meshes with 8-level octree.
    } 
    \vspace{-0.5cm}

    \label{fig:objaverse}
\end{figure*}

We present the qualitative results of mesh generation conditioned on input point cloud in Figure~\ref{fig:objaverse}, comparing our method with MeshAnything~\cite{chen2024meshanything} and MeshAnythingV2~\cite{chen2024meshanythingv2} as the baselines. As shown, our method demonstrates a notable improvement to generate meshes with higher face counts and refined details.

\begin{figure*}[!ht]
\vspace{-2mm}
    \centering
    \setlength{\tabcolsep}{1pt}
    \begin{tabular}{c|cc|cc|cc} 
        \textbf{GT Mesh} & \multicolumn{2}{c|}{\textbf{MeshAnything}~\cite{chen2024meshanything}} & \multicolumn{2}{c|}{\textbf{MeshAnythingV2}~\cite{chen2024meshanythingv2}} & \multicolumn{2}{c}{\textbf{Ours}} \\ 


        \centered{\includegraphics[width=\www\linewidth]{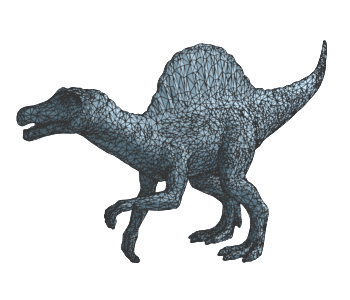}}
        & \centered{\includegraphics[width=\www\linewidth]{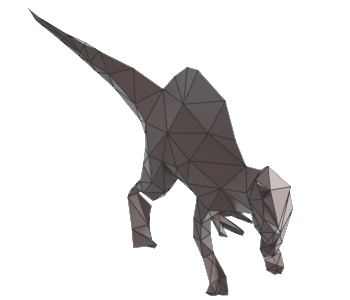}}
        & \centered{\includegraphics[width=\www\linewidth]{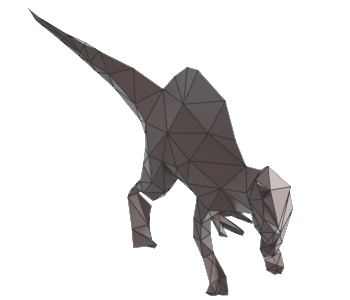}}
        & \centered{\includegraphics[width=\www\linewidth]{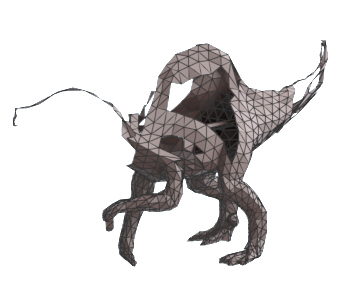}}
        & \centered{\includegraphics[width=\www\linewidth]{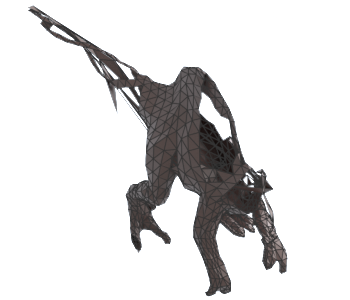}}
        & \centered{\includegraphics[width=\www\linewidth]{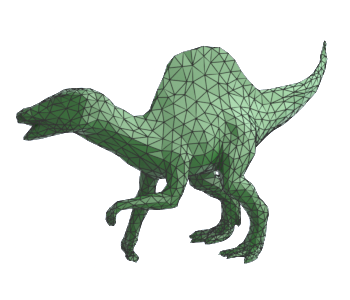}}
        & \centered{\includegraphics[width=\www\linewidth]{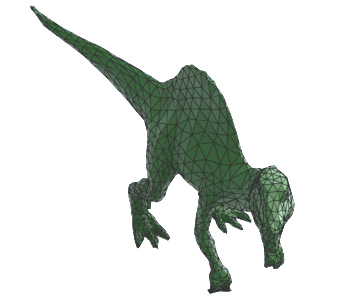}} \\

        \#Face = 14430 & \multicolumn{2}{c|}{\#Face = 506} & \multicolumn{2}{c|}{\#Face = 2219} & \multicolumn{2}{c}{\#Face = 2877} \\ 


        \centered{\includegraphics[width=\www\linewidth]{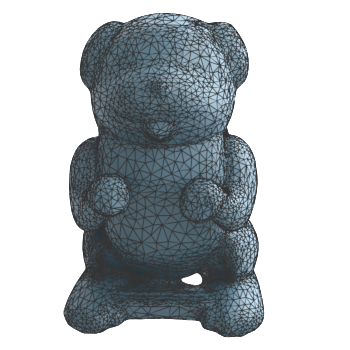}}
        & \centered{\includegraphics[width=\www\linewidth]{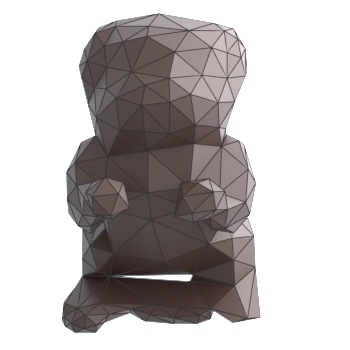}}
        & \centered{\includegraphics[width=\www\linewidth]{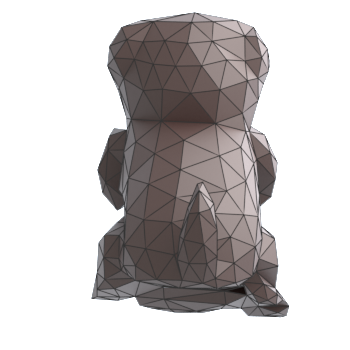}}
        & \centered{\includegraphics[width=\www\linewidth]{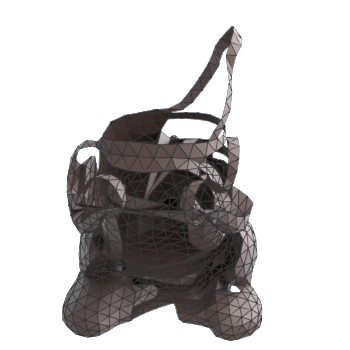}}
        & \centered{\includegraphics[width=\www\linewidth]{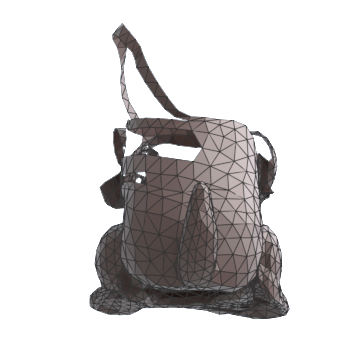}}
        & \centered{\includegraphics[width=\www\linewidth]{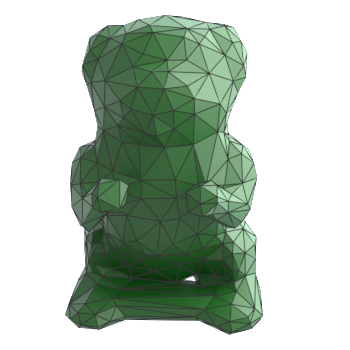}}
        & \centered{\includegraphics[width=\www\linewidth]{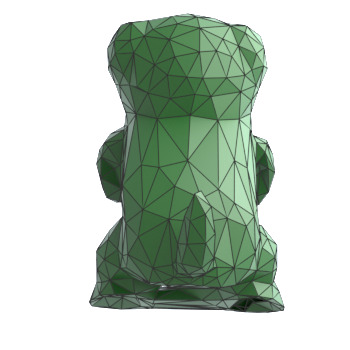}} \\

    \#Face = 15672 & \multicolumn{2}{c|}{\#Face = 634} & \multicolumn{2}{c|}{\#Face = 2161} & \multicolumn{2}{c}{\#Face = 939} \\
        
    \end{tabular} \vspace{-0.2cm}
    \caption{
    \textbf{Qualitative comparison on GSO dataset~\citep{downs2022google}.}
    } 
    \vspace{-0.4cm}
    \label{fig:gso}
\end{figure*}

\subsubsection{Quantitative Results}

We use the following metrics to assess the quality of the generated meshes:

\begin{itemize}
    \item \textbf{Chamfer Distance (CD).} It provides an indication of geometric accuracy by measuring the average closest-point distance between points sampled from the source and reference meshes, computed bidirectionally. Lower values of CD indicate better geometric accuracy.
    \item \textbf{Normal Consistency (NC).} It evaluates the surface orientation alignment between the source and reference meshes. Specifically, it measures the cosine similarity between the normals of each face in the source mesh and the nearest face in the reference mesh, computed bidirectionally. Higher NC values signify better normal alignment. We also report absolute Normal Consistency ($|$NC$|$), which disregards the sign, focusing solely on the magnitude of similarity. The mathematical details are provided in the supplementary materials.
\end{itemize}

\boldparagraph{Objaverse evaluation set} We ran the evaluation on 200 samples in our validation set with 1 generation seed for each model. The quantitative results presented in Table~\ref{tab:objaverse_quant}, indicate that our model generates meshes more faithful to the ground truth meshes compared to the baselines. The NC values of MeshAnything~\citep{chen2024meshanything} and MeshAnythingV2~\citep{chen2024meshanythingv2} are relatively low due to flipped normals, which leads to inconsistencies in the sign of the normals. In contrast, our method generates meshes with consistent normal orientation.

\begin{table}[!h]
\centering
\normalsize  
\begin{tabular}{lS[table-format=1.4]S[table-format=1.4]S[table-format=1.4]}
\toprule
\textbf{Model} & \textbf{CD$\downarrow$} & \textbf{NC$\uparrow$} & \textbf{$|$NC$|\uparrow$} \\ \midrule
MeshAnything~\citep{chen2024meshanything}  & 0.0115      & 0.223       & 0.853           \\
MeshAnythingV2~\citep{chen2024meshanythingv2} & 0.0102      & 0.167      & 0.843           \\
Ours           & \textbf{0.0070} & \textbf{0.798} & \textbf{0.880} \\ \bottomrule
\end{tabular} \vspace{-0.2cm}
\caption{\textbf{Quantitative results on Objaverse evaluation set.}}
\label{tab:objaverse_quant} 
\end{table}

\boldparagraph{Tokenization effectiveness} The previous trained models and ours differ in dataset composition, point cloud conditioning, and training settings. To accurately evaluate the effectiveness of different tokenization methods, we conduct a controlled experiment on the subset of our filtered dataset of 24k samples with $\leq$500 faces. All factors except for the tokenizers are kept the same: 1). 16-layer Transformers with 768 hidden dimension and positional encoding. 2). Point cloud condition of 2048 tokens. 3). Training until 40k steps with an effective batch size of 128 and data augmentation. 

Results on a test set of 200 samples, shown in Table~\ref{tab:tokenization_comparison}, strongly indicate our method provides highly effective inductive bias. We use PivotMesh's~\citep{weng2024pivotmesh} pretrained VQ-VAE since MeshAnything does not release their VQ-VAE encoder and noise resistant decoder fine-tuning code.

\begin{table}[!h]
\setlength{\tabcolsep}{3pt} 
\centering
\normalsize  
\begin{tabular}{lS[table-format=1.4]S[table-format=1.4]S[table-format=1.3]p{2cm}}
\toprule
\textbf{Tokenizer} & \textbf{CD$\downarrow$} & \textbf{NC$\uparrow$} & \textbf{$|$NC$|\uparrow$} & \textbf{Sequence Length$\downarrow$} \\ \midrule
Naive~\citep{chen2024meshxl}  & {0.0376} & {0.639} & {0.822} & $9N_f$ \\
VQ-VAE~\citep{weng2024pivotmesh}  & {0.0352} & {0.673} & {0.815} & $6N_f$ \\
AMT~\citep{chen2024meshanythingv2} & {0.0327} & {-0.069} & {0.768} & $\pm 4 N_f$ \\
Ours           & \textbf{0.0100} & \textbf{0.734} & \textbf{0.874} & $\bm{2N_f+2N_c}$ \\ \bottomrule
\end{tabular} \vspace{-0.2cm}
\caption{\textbf{Quantitative results on our controlled experiment.}  $N_f$ and $N_c$ are the number of triangular faces and connected components, respectively. }
\label{tab:tokenization_comparison} \vspace{-0.4cm}
\end{table}

\subsection{Results on GSO Dataset}

We further conduct a quantitative evaluation on GSO dataset~\citep{downs2022google}, a dataset of real-world 3D scans to test the generalization capabilities of each model. In this experiment, we observe that the baseline models and ours are sensitive to the input point cloud, with results varying large on different sampling seeds. To mitigate this variability, we generate five samples for each model and select the mesh with the lowest Chamfer Distance for evaluation. 

For our model, we sample point clouds directly from the meshes provided in the GSO dataset. However, direct sampling from these high-resolution meshes often results in reconstructions with extremely small faces. To address this, we decimate the meshes to five target faces ranging from 1000 to 2500 faces and then perform uniform sampling on these decimated versions. For MeshAnything and MeshAnythingV2, we sample the input point clouds from Marching Cubes meshes derived from both the original high-resolution and decimated (1500 faces) meshes using an 8-level octree.

\begin{figure}[!h]
    \centering
    \setlength{\tabcolsep}{0.7pt}
    \begin{tabular}{ccc} 
        \small{\textbf{Original}} & \small{\textbf{Decimated}} & \small{\textbf{Ours}} \\ 
        \centered{\includegraphics[width=0.28\linewidth]{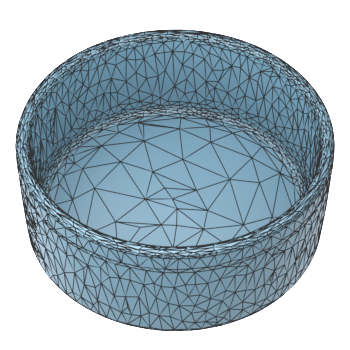}}
        & \centered{\includegraphics[width=0.28\linewidth]{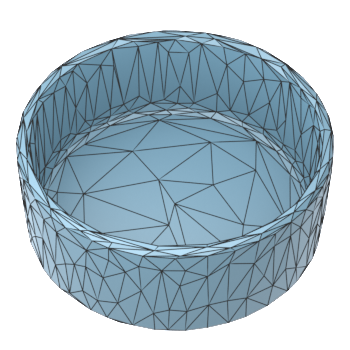}}
        & \centered{\includegraphics[width=0.28\linewidth]{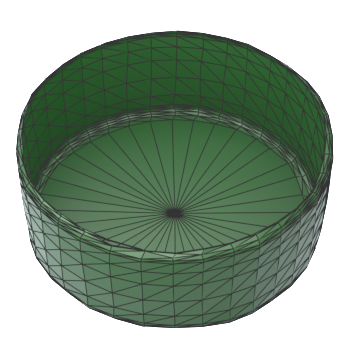}} \\
        
    \end{tabular}\vspace{-0.4cm}
    \caption{
    \textbf{Comparison between the decimated mesh and our output.} Our model is capable of generating meshes with the topology of those created by 3D artists.
    } 
    \vspace{-0.4cm}
    \label{fig:decimated}
\end{figure}

As shown in Table~\ref{tab:model_comparison}, our method achieves the best results across the evaluated metrics. Figure~\ref{fig:gso} presents qualitative comparisons, demonstrating that our model generates meshes more consistent with the input than the baselines. Additionally, Figure~\ref{fig:decimated} compares our output with the decimated mesh, highlighting that our model can generate meshes with the topology of those created by 3D artists.

\begin{table}[!h]
\centering
\normalsize
\begin{tabular}{lS[table-format=1.4]S[table-format=1.4]S[table-format=1.4]}
\toprule
\textbf{Model} & \textbf{CD$\downarrow$} & \textbf{NC$\uparrow$} & \textbf{$|$NC$|\uparrow$} \\ \midrule
MeshAnything~\cite{chen2024meshanything}   & 0.0105      & 0.453       & 0.869           \\
MeshAnythingV2~\cite{chen2024meshanythingv2} & 0.0116      & 0.3269      & 0.865           \\
Ours           & \textbf{0.0077} & \textbf{0.842} & \textbf{0.897} \\ \bottomrule
\end{tabular} \vspace{-0.2cm}
\caption{\textbf{Quantitative results on GSO dataset.}}
\label{tab:model_comparison}
\end{table}\vspace{-0.4cm}

\subsection{Ablation Study}\label{sec:ablation}

\begin{figure}[!h]
\vspace{-0.2cm}
    \centering
    \setlength{\tabcolsep}{0.7pt}
    \begin{tabular}{ccc} 
        \small{\textbf{GT Mesh}} &\small{\textbf{Simultaneous}} & \small{\textbf{Hierarchical}} \\ 
        \centered{\includegraphics[width=0.28\linewidth]{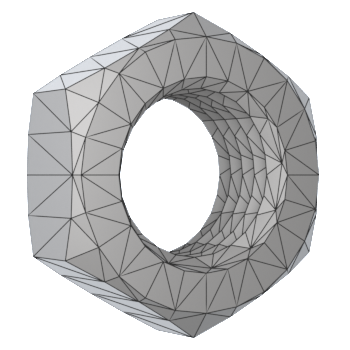}}
        & \centered{\includegraphics[width=0.28\linewidth]{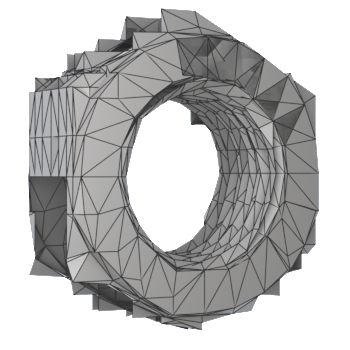}}
        & \centered{\includegraphics[width=0.28\linewidth]{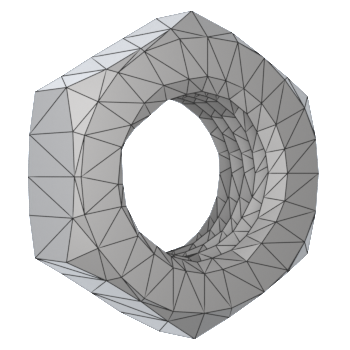}} \\
        
    \end{tabular} \vspace{-0.35cm}
    \caption{
    \textbf{MLP head ablation.} Our hierarchical MLP maintains the sequential nature of the x-y-z coordinates prediction that results in easier sampling compared to simultaenous prediction.
    } 
    \vspace{-0.4cm}
    \label{fig:head_ablation}
\end{figure}

\boldparagraph{Decoder MLP head} We compare TreeMeshGPT trained with MLP heads that predict the x-y-z coordinates simultaneously and our proposed hierarchical MLP. Predicting the coordinate simultaneously leads to challenging sampling that results in noisy and more incomplete meshes, as shown in Figure~\ref{fig:head_ablation}. Running evaluation on the 200 validation samples used in Table~\ref{tab:objaverse_quant} with this head yields CD = $0.0114$, NC = $0.724$, $|$NC$|$ = $0.847$.

\begin{figure}[h]
    \centering
    \includegraphics[width=0.47\textwidth]{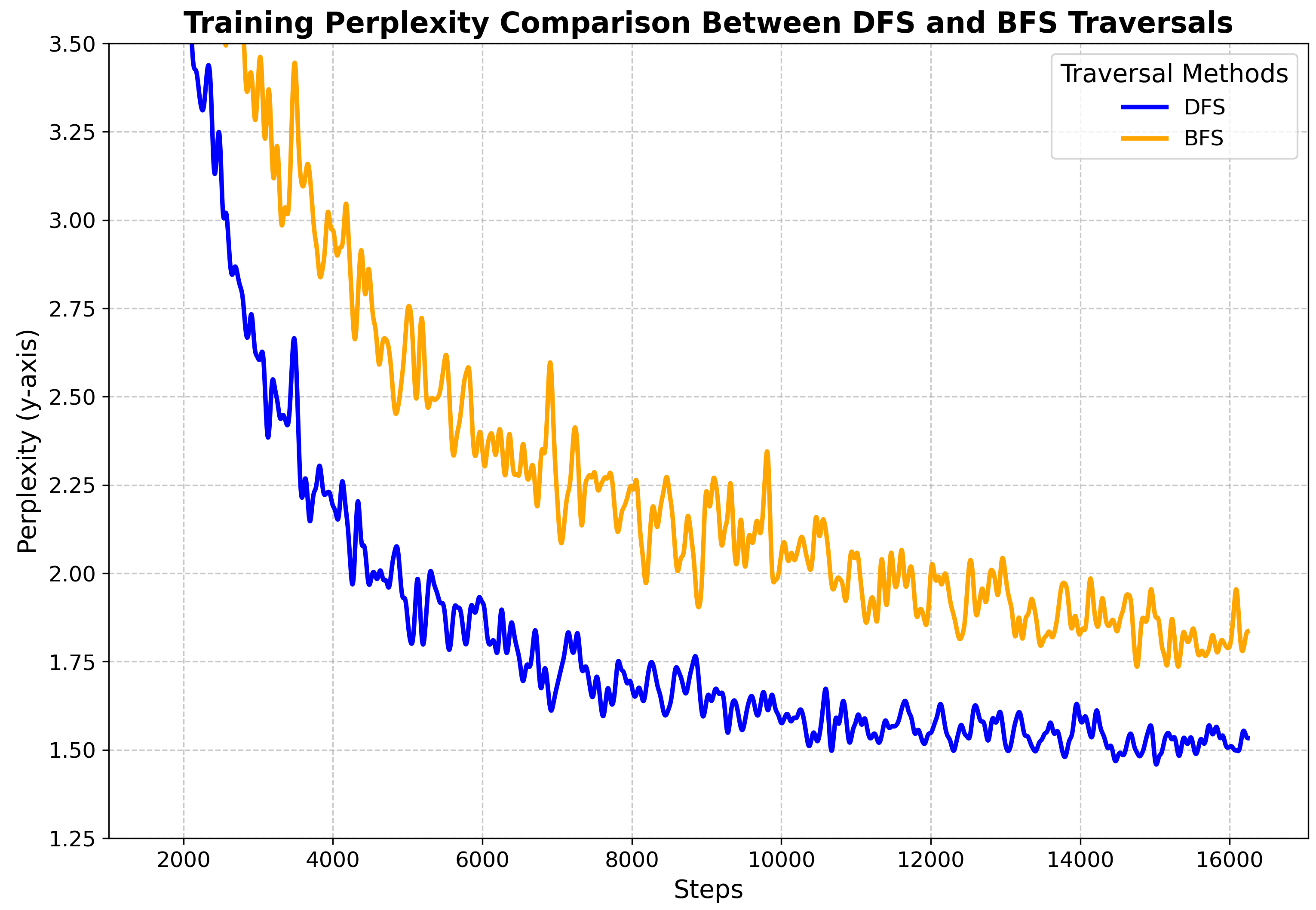} 
    \vspace{-0.3cm}\caption{\textbf{Training perplexity comparison Between BFS and DFS traversals.} DFS traversal shows a better training perplexity compared to BFS. Shown in the plot is the perplexity for $y-$axis vertex coodinate.}
    \label{fig:bfs_dfs_comparison}
\end{figure}

\boldparagraph{Tree traversal} We conduct an ablation study on smaller Transformer architecture comparing DFS and breadth-first-search (BFS) traversals in forming input-output sequences for our Autoregressive Tree Sequencing. As shown in the training perplexity plot in Figure~\ref{fig:bfs_dfs_comparison}, DFS traversal enables more efficient Transformer training. This improvement likely stems from the stronger local dependencies introduced by DFS, where each step is more predictable based on its immediate predecessors. In contrast, BFS traversal often introduces dependencies between steps that are spatially or structurally distant, and thus complicating the learning process.

\vspace{-0.2cm}
\vspace{-0.5cm}

\section{Conclusion}

We present \OURS{}, an autoregressive Transformer designed to generate high-quality artistic meshes aligned with input point clouds. \OURS{} incorporates a novel Autoregressive Tree Sequencing technique instead of the conventional next-token prediction. Our Autoregressive Tree Sequencing represents each face with two tokens, enabling a 7-bit discretization model that can generate up to 5,500 triangular faces with 2,048 point cloud latent tokens. Experiments show that \OURS{} can generate meshes with higher quality compared to the previous methods.

\boldparagraph{Limitations} Our model has a similar failure mode to the previous methods that the success rate decreases as the sequence length increases. Additionally, while our model has an improved capacity to generate meshes with higher face counts, challenges persist in enforcing an optimal mesh topology.

\noindent{\textbf{Acknowledgment.}} This research is supported by the National Research Foundation (NRF) Singapore, under its NRF-Investigatorship Programme (Award ID. NRF-NRFI09-0008).

{
    \small
    \bibliographystyle{ieeenat_fullname}
    \bibliography{main}
}

\newpage
\appendix




%

\makeatletter
\setlength{\@fptop}{0pt}
\makeatother


\def\paperID{14517} 
\def\confName{CVPR}
\def\confYear{2025}

\onecolumn
\section*{\Large \centering Supplementary Material for\\ \OURS: Artistic Mesh Generation with Autoregressive Tree Sequencing}
\vspace{7mm}


In this supplementary document, we provide the implementation details of our network's MLP heads in Section~\ref{sec:implementation}. Then, we provide the mathematical details of the Normal Consistency metrics in Section~\ref{sec:metric}. We then demonstrate the capability of \OURS to generate artistic meshes from text prompts through a multi-step process in Section~\ref{sec:genie}. Finally, we present our 9-bit model supporting the generation of artistic meshes with up to 11,000 faces in Section~\ref{sec:9bit}.

\vspace{0.5cm}

\section{Vertex Prediction Heads}\label{sec:implementation}


\begin{figure}[h]
  \centering
  \includegraphics[width=0.8\textwidth]{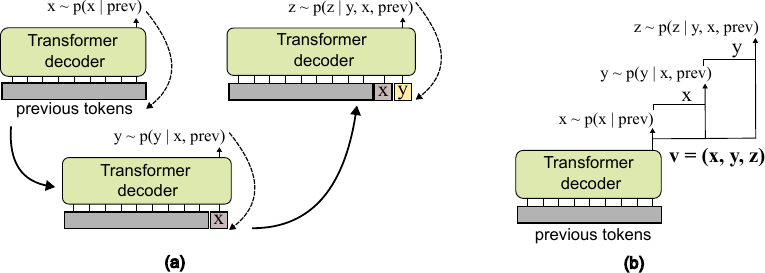}
  \caption{\textbf{Sequential vertex prediction.} a). Next-token prediction. b). Our hierarchical MLP heads.}
  \label{fig:hierMLP}
\end{figure}

To mimic the sequential nature in the prediction of vertex’s \( x \)-, \( y \)-, and \( z \)-coordinates
in next-token prediction Transformer (Figure 1a), we adopt hierarchical MLP heads (Figure 1b). Our hierarchical MLP heads contain three stages to the generate each vertex’s \( x \)-, \( y \)-, and \( z \)-coordinates, where each coordinate is predicted sequentially based on the previous ones. In the first stage of the hierarchical MLP, represented by \( g_{\theta 1} \), the initial coordinate (e.g., \( x \)-coordinate) of the vertex is predicted based on the latent code \( \mathbf{c} \in \mathbb{R}^d \) from the Transformer decoder:

\begin{equation}
    x \sim p(x \mid \text{prev}) = g_{\theta 1}(\mathbf{c}).
\end{equation} Here, "prev" denotes all previously generated tokens, and \( \mathbf{c}\) is the latent code output by the Transformer decoder, which encapsulates information from these prior tokens. Next, the \( y \)-coordinate is predicted in the second stage of the MLP, represented by \( g_{\theta 2} \):

\begin{equation}
    y \sim p(y \mid x, \text{prev}) = g_{\theta 2}(\text{E}_x(x), \mathbf{c}),
\end{equation} where \(\text{E}_{*} \in \mathbb{R}^d \) denotes the learnable embeddings for the discretized coordinates of an axis and \(*\) can represent \(x\), \(y\), or \(z\). For example, \(\text{E}_x(x)\) represents the embedding of the discretized \(x\)-coordinate, and similarly, \(\text{E}_y(y)\) and \(\text{E}_z(z)\) denote the embeddings of the discretized \(y\)- and \(z\)-coordinates, respectively. This second stage conditions on both the latent code \( \mathbf{c} \) and the discretized coordinate embedding \(\text{E}_{x}\). Finally, the \( z \)-coordinate is predicted in the third stage of the MLP, \( g_{\theta 3} \), which takes as input the latent code \( \mathbf{c} \) along with the embeddings of both previously predicted coordinates, \(\text{E}_x(x)\) and \(\text{E}_y(y)\):

\begin{equation}
    z \sim p(z \mid y, x, \text{prev}) = g_{\theta 3}(\text{E}_y(y), \text{E}_x(x), \mathbf{c}).
\end{equation}  In each stage, the input to the MLP \( g_{\theta} \) consists of the concatenation of the latent code \( \mathbf{c} \) and the corresponding embeddings \( \text{E}_{*} \). 

In our experiments with the Objaverse dataset, where the \( z \)-axis represents the height axis, we predict the \( z \)-coordinate first, followed by the \( y \)-coordinate and then the \( x \)-coordinate. Additional \texttt{[STOP]} and \texttt{[EOS]} labels are included in the class selection for the \( z \)-coordinate. During training, the loss functions for the \( y \)- and \( x \)-coordinates are applied only when the ground truth \( z \)-coordinate is not one of these additional labels. Also, teacher-forcing is employed to supervise the \( y \)- and \( x \)-coordinates by conditioning with the embeddings of the preceding ground truth coordinates.

\vspace{0.5cm}

\section{Normal Consistency Metrics}\label{sec:metric}

This section details the calculation of our normal consistency metrics. Let \( \mathcal{M}_s \) and \( \mathcal{M}_r \) denote the source and reference meshes, respectively, where each consists of triangular faces. The centroid \( \mathbf{c}^s_i \) of the \(i\)-th face in the source mesh \( \mathcal{M}_s \) is given by:

\begin{align*}
\mathbf{c}^s_i = \frac{\mathbf{v}^s_{i1} + \mathbf{v}^s_{i2} + \mathbf{v}^s_{i3}}{3},
\end{align*} where \( \mathbf{v}^s_{i1}, \mathbf{v}^s_{i2}, \mathbf{v}^s_{i3} \) are the vertices of the \(i\)-th triangular face of \( \mathcal{M}_s \). For each centroid \( \mathbf{c}^s_i \), we find the closest face \( j \) on the reference mesh \( \mathcal{M}_r \) using the shortest point-to-face distance:

\begin{align*}
j = \arg\min_{k \in \mathcal{M}_r} d(\mathbf{c}^s_i, F^r_k),
\end{align*} where \( F^r_k \) is the \(k\)-th face in \( \mathcal{M}_r \) and \( d(\mathbf{c}^s_i, F^r_k) \) represents the shortest distance from the point \( \mathbf{c}^s_i \) to the face \( F^r_k \). The cosine similarity between the normals of the \(i\)-th face in the source mesh (\( \mathbf{n}_i^s \)) and the closest face (\( \mathbf{n}_j^r \)) in the reference mesh is then computed as:

\begin{align*}
\text{Sim}_{i \to j}(\mathbf{n}^s, \mathbf{n}^r) = \frac{\mathbf{n}_i^s \cdot \mathbf{n}_j^r}{\|\mathbf{n}_i^s\| \|\mathbf{n}_j^r\|}.
\end{align*}

This process is repeated bidirectionally. For the reverse direction, the centroid \( \mathbf{c}^r_k \) of the \(k\)-th face in \( \mathcal{M}_r \) is computed to find the corresponding closest face \( l \) in \( \mathcal{M}_s \). The Normal Consistency (NC) metric is the average cosine similarity across all face pairs in both directions:

\begin{equation}
\text{NC} = \frac{1}{2|\mathcal{M}_s|} \sum_{i \in \mathcal{M}_s} \text{Sim}_{i \to j}(\mathbf{n}^s, \mathbf{n}^r) + \frac{1}{2|\mathcal{M}_r|} \sum_{k \in \mathcal{M}_r} \text{Sim}_{k \to l}(\mathbf{n}^r, \mathbf{n}^s),
\end{equation} where \( |\mathcal{M}_s| \) and \( |\mathcal{M}_r| \) are the numbers of faces in the source and reference meshes, respectively and \( l =  \arg\min_{i \in \mathcal{M}_s} d(\mathbf{c}^r_k, F^s_i) \). The absolute version ($|$NC$|$) that omits the flipping direction is then given as:

\begin{equation}
|\text{NC}| = \frac{1}{2|\mathcal{M}_s|} \sum_{i \in \mathcal{M}_s} |\text{Sim}_{i \to j}(\mathbf{n}^s, \mathbf{n}^r)| + \frac{1}{2|\mathcal{M}_r|} \sum_{k \in \mathcal{M}_r} |\text{Sim}_{k \to l}(\mathbf{n}^r, \mathbf{n}^s)|.
\end{equation} 

\vspace{0.5cm}
\section{Generating Artistic Meshes from Text Prompts}\label{sec:genie}

\begin{figure}[h]
    \centering
    \includegraphics[width=0.97\textwidth]{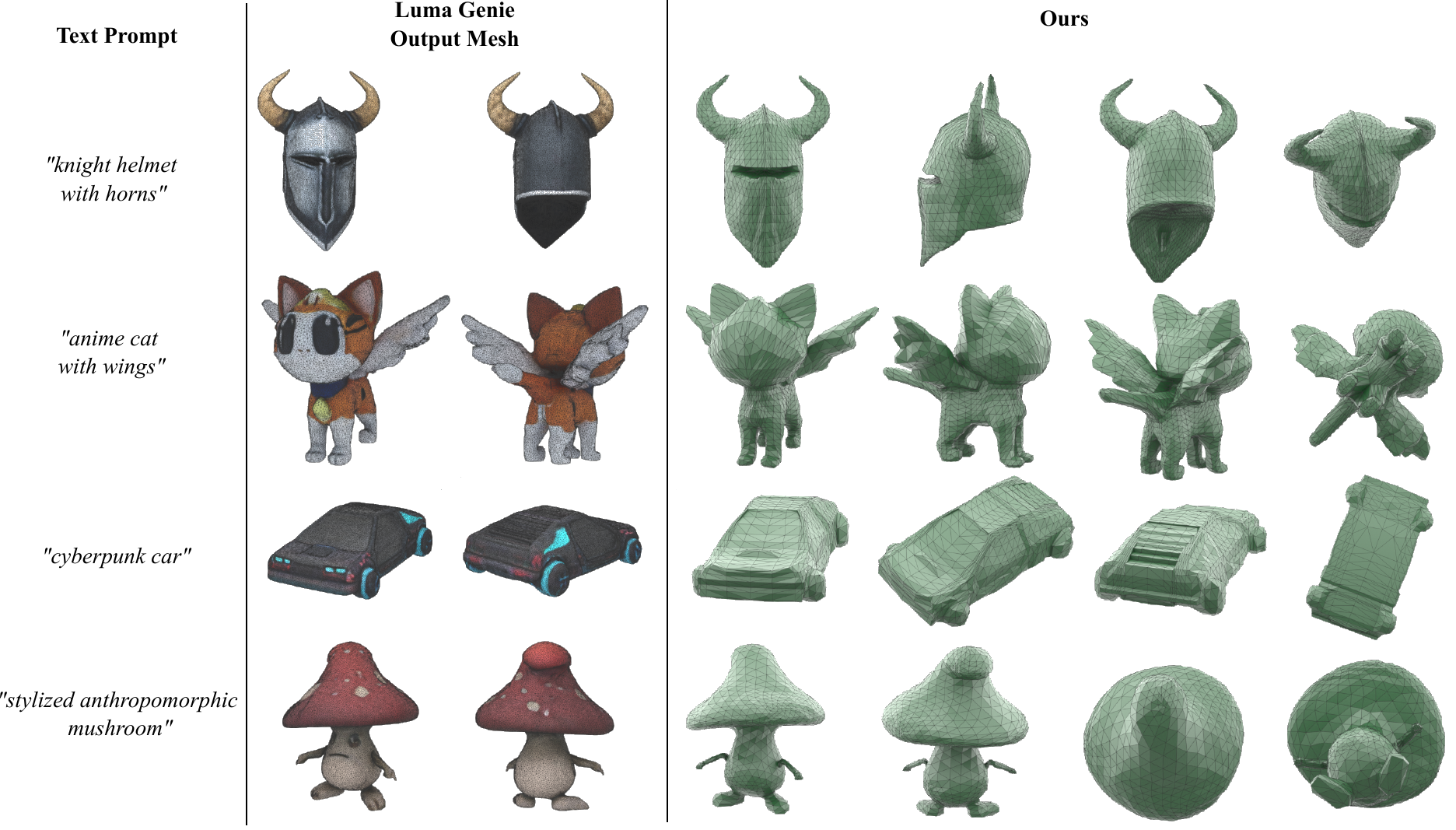} 
    \caption{\textbf{Multi-step text-to-artistic mesh generation.} Given a text prompt, we first generate a dense mesh using the Luma AI Genie model. This dense mesh, typically containing around 50,000 triangles, is then decimated. A point cloud is sampled from the decimated mesh and serves as the input condition for \OURS, which generates the final artistic mesh.}
    \label{fig:genie}
\end{figure}

We demonstrate the capability of our model to generate artistic meshes from text prompts through a multi-step process, shown in Figure~\ref{fig:genie}. We utilize the Luma AI Genie\footnote{\url{https://lumalabs.ai/genie}} text-to-3D model to generate dense meshes from text prompts. These meshes are typically over-tessellated, containing around 50,000 small triangles that make them unsuitable for downstream applications. To generate artistic meshes, we first apply decimation to the dense meshes. Next, we sample point clouds from the decimated meshes and use them as input conditions of \OURS. 

\vspace{0.5cm}

\vspace{0.5cm}
\section{9-bit Model Supporting 10K+ Faces}\label{sec:9bit}

In our model training with 7-bit discretization, we performed the discretization to the normalized manifold Objaverse meshes, removed the duplicate triangles, and chose meshes with $\leq$5.5k faces as we found significant amount of these discretized meshes with $>$5.5k faces contain small triangles that collapse or merge, thus violating the manifold condition required for our sequencing approach. 

\begin{figure}[p]
    \centering
    \includegraphics[width=0.97\textwidth]{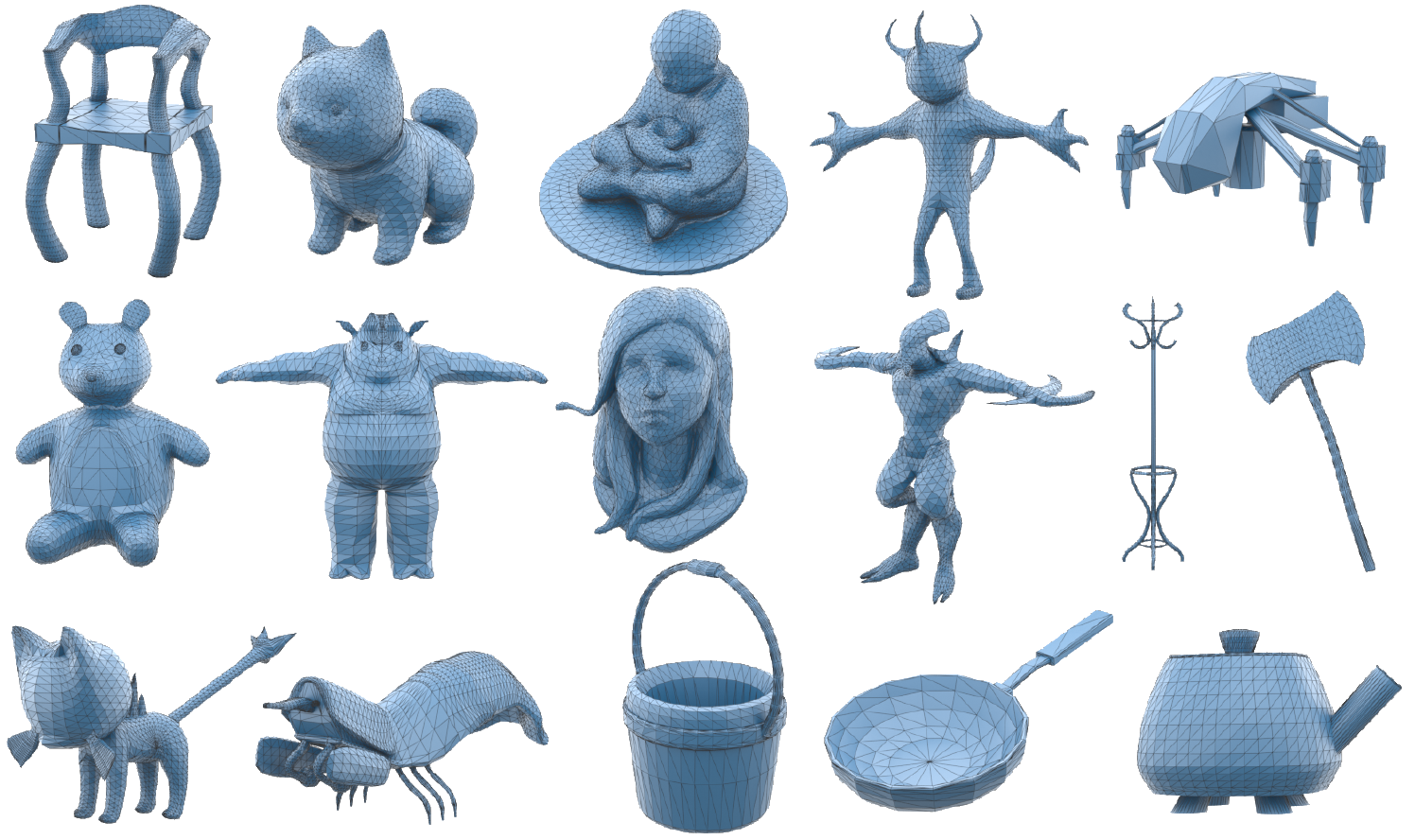} 
    \caption{\textbf{Qualitative results of our 9-bit model.} The generated meshes contain up to 11,000 faces, demonstrating improved surface smoothness and finer details compared to the 7-bit model. Inputs are point clouds sampled from Objaverse meshes.}
    \label{fig:9bit}
\end{figure}

These triangles collapses/merges occur less with finer discretization and we further train \OURS{} with 9-bit discretization. Our 9-bit model supports the generation of up to 11,000 faces, taking 25 days of training with $8\times$ A100-80GB GPUs. Some of the qualitative results are shown in Figure~\ref{fig:9bit}. Compared to the 7-bit model, our 9-bit model can generate artistic meshes with smoother surfaces, finer details, and higher number of faces.

\end{document}